\documentclass[a4paper, amsfonts, amssymb, amsmath, reprint, showkeys, nofootinbib, twoside]{revtex4-1}
\usepackage[english]{babel}
\usepackage[utf8]{inputenc}
\usepackage[colorinlistoftodos, color=green!40, prependcaption]{todonotes}
\begin{document}

\title{Stirring the quantum vacuum: Angular Casimir Momentum of a Landau Charge}
\author{B.A. van Tiggelen}
    \email[Correspondence email address: ]{bart.van-tiggelen@grenoble.cnrs.fr}
    \affiliation{Univ. Grenoble Alpes, CNRS, LPMMC, 38000 Grenoble, France}

\date{\today} 

\begin{abstract}
 We consider the angular momentum of a charge $q$ rotating in a homogeneous magnetic field and study the role of the electromagnetic quantum vacuum.
Its orbital angular momentum is caused by the recoil of energetic vacuum photons that grows as $n^2$,
 i.e. faster than the kinetic angular momentum $-2n\hbar$ of a Landau level.
\end{abstract}


\maketitle

\section{Introduction}
The radiation of the electromagnetic (EM) quantum vacuum is perfectly isotropic and possesses an
energy density $\hbar \omega d\omega/\pi^3c_0^3  $ in the frequency interval $d\omega$.
This statement is Lorentz-invariant \cite{milonni}. The mystery of its UV-divergence is still one of the major challenges in physics,
but poses in general no problems to calculate the Casimir force between dielectric or metallic objects \cite{Casi}.
Due to its perfect isotropy, energy current and momentum density of the quantum vacuum vanish.

This is no longer true  when the  vacuum  interacts with matter. In a pioneering work \cite{Feigel} Feigel predicted
''Casimir" momentum in bi-anisotropic materials. Although
the Poynting vector of the quantum vacuum  still vanishes \cite{BartEPJD},
a momentum density emerges that suffers from a
divergence at high energies. Relativistic photons were already known to be relevant for the Lamb shift \cite{milonni},
in contrast to Casimir-Polder forces \cite{Casi}, caused by low-energy vacuum photons.
In a microscopic bi-anisotropic quantum model,
the divergency of EM momentum
is removed  by mass renormalization \cite{kawka}, resulting in a Casimir momentum of the order
of $\alpha^2$ times the classical Abraham momentum \cite{loudon,Geert1}. It was also demonstrated
for a chiral quantum particle in a magnetic field \cite{Donaire,Donaire2}.

EM angular momentum is of great recent interest \cite{EMAM} and in macroscopic media even controversial \cite{AMdiel},
since matter and radiation
are hard to disentangle.
In Ref.~\cite{astrid} Casimir-Polder torques were discussed to stir vortices in superfluids.
How does ''Casimir" angular momentum emerge on a microscopic scale?
The simplest model for which this question can be answered is the well-known cyclotron
problem of a non-relativistic charge $q$ without spin rotating in a uniform magnetic field $\mathbf{B}_0$, in quantum-mechanics better known as
the Landau problem.
In the Coulomb gauge and in Gaussian units, the vector potential is  $\mathbf{A }_0(\mathbf{r},t)=  \mathbf{B}_0(t) \times\mathbf{r} /2$.
With the quantum vacuum, the non-relativistic Hamiltonian reads,
 \begin{equation}\label{HQV}
    H = \frac{1}{2\mu}\left(\mathbf{p}- \frac{q}{c_0}\mathbf{A}_0(\mathbf{r},t) - \frac{q}{c_0}\mathbf{A}(\mathbf{r}) \right)^2 + \sum_{{k}\mathbf{\Pi} } \hbar \omega_k  a^\dag_{k\mathbf{\Pi}}a_{k\mathbf{\Pi}}
\end{equation}
Here, $\mathbf{p}$ the canonical momentum satisfying $[r_i,p_j] = i\hbar \delta_{ij}$.
It is customary to choose the $z$-axis along the direction of $q\mathbf{B}_0$ and to introduce the cyclotron frequency $\omega_c= qB_0/\mu c_0 $.
It is convenient to express the vector potential of the quantum vacuum in terms of the spherical vector harmonics $\Phi_{\mathbf{\Pi}}(\mathbf{\hat{k}})$ defined on the unit sphere in
reciprocal space  \cite{Cohen},
\begin{eqnarray}\label{AQV}
    \mathbf{A}(\mathbf{r}) = \sqrt{2\pi \hbar c_0} \sum_{\mathbf{k}\mathbf{\Pi} } &&\frac{1}{k^{3/2}} \left(  a_{k\mathbf{\Pi}}\exp(-i\mathbf{k}\cdot \mathbf{r}) \Phi_{\mathbf{\Pi}}(\mathbf{\hat{k}}) \right. \nonumber \\
&&  + \left.   a^\dag_{k\mathbf{\Pi}}\exp(i\mathbf{k}\cdot \mathbf{r}) \bar{{\Phi}}_{\mathbf\Pi}(\mathbf{\hat{k}}) \right)
\end{eqnarray}
The continuum limit has been taken, $\sum_\mathbf{k} \equiv \int d^3 \mathbf{k}/(2\pi)^3$ and
$[a_{k\mathbf{\Pi}}, a^\dag_{k'\mathbf{\Pi'}}]  = \delta_{kk'}\delta_{\mathbf{\Pi},\mathbf{\Pi}'}$.
The vector index $\mathbf{\Pi} =\{J,M,p\} $ summarizes the 3 discrete quantum numbers of total EM angular momentum
($J$), its $z$-component ($M$) and two transverse polarizations $p$, the longitudinal vector harmonic being excluded by the
Coulomb gauge. Without the quantum vacuum, the kinetic momentum of the charge is $\mathbf{p}^K= \mu d\mathbf{r}/dt = \mu [H_0,\mathbf{r}]/i\hbar  =
 \mathbf{p}- q\mathbf{A}_0(\mathbf{r},t)/c_0$. The canonical angular momentum $l_z=(\mathbf{r}\times \mathbf{p})_z$ commutes
 with $H_0$ and has eigenvalues $(|m|-n) \hbar$. The Hamiltonian $H_0$ has the eigenstates  $|n,m,k_z\rangle $ of a 2D harmonic oscillator
 whose energy levels  $E_n= (n+\frac{1}{2})\hbar\omega_c + \hbar^2k_z^2/2\mu $ are independent of $m$ due to gauge invariance \cite{Goerbig}.

In general,  total angular momentum $\mathcal{L}_z$ is conserved, equal to the sum of the canonical angular momentum $l_z$ of the charge and the
 angular momentum of the  transverse EM field $J_z$.
 The angular kinetic momentum operator of the charge is not conserved and can be written as,
\begin{equation}\label{LLkinQV}
    l_z^{\mathrm{K}}= \mathcal{L}_z  -  J_z -
    \frac{q}{c_0}(\mathbf{r}\times \mathbf{A})_z -l^L
 \end{equation}
The ''Lenz" term $l^{L} = \mu \omega_c \rho^2/2$ is responsible for the electromotive force
in classical electrodynamics, recently discussed
quantum-mechanically \cite{Barnettl} as well  as key element in the EM momentum controversy \cite{AM}.  Without the quantum vacuum, the
kinetic angular momentum of a Landau state is given by
$ \langle n,m| l_z^{K}|n,m\rangle = -(2n+1) \hbar = -2E_n(\mathrm{rot})/\omega_c$, again independent on $m$. When the quantum vacuum  is included,
$\langle l_z^{K} \rangle $
will decay radiatively, but slow enough for $\langle l_z^{K} \rangle $ to be well-defined and modified.

We will use time-dependent perturbation theory to calculate the different contributions of the quantum vacuum to
$l_z^{\mathrm{K}}$ for an excited  Landau state, and all proportional to $qB_0(t)$ for slow adiabatic changes.
We identify the interaction  $W = -(q/\mu c_0) \mathbf{p}^K\cdot \mathbf{A}(\mathbf{r}) $ between rotating charge and quantum vacuum,
and imagine to switch it on slowly like  $W(t)= W \exp(\epsilon t/\hbar )$ at $t_0 \rightarrow -\infty$,
when the total wave function is assumed to be in the pure state $|N\rangle = | n,m,k_z=0 \rangle \otimes |\{0\}\rangle$ \cite{loudonboek}. In this processus $\mathcal{L}_z$ is conserved and equal to its initial value $(|m|-n)\hbar$.
The index $N'$ refers to all possible product states $\{n,m, k_z\} \otimes \{n_{k\mathbf{\Pi}}\}$ of charge plus transverse photons. Explicit reference will be made
nor to the highly degenerated levels $m$, neither to the momentum $k_z$ of the charge along the magnetic field. The
degeneration of the $m$-levels is protected by gauge invariance,
and the impact of photon recoil on the longitudinal displacement is negligible.
Since  $W_{NN}=0$, the wave function at $t\geq t_0$ is perturbed as,
\begin{eqnarray}\label{perturb}
 \nonumber  |\Psi_N (t)\rangle  &=&  \exp\left( -\frac{i}{\hbar}\int^t_{t_0} dt' \left[E_N+ \Delta E_N(t') \right]\right) |N\rangle \\
 \nonumber    +&& \sum'_{N'}  |N'\rangle\frac{W_{N'N}(t)  }{E_N-E_{N'} + i\epsilon} + \cdots
\end{eqnarray}
with $\Delta E_N(t) = \sum'_{N'}  W_{NN'}(t)W_{N'N}(t)/ (E_N-E_{N'} +i\epsilon)  $ the second-order perturbation of the energy level $E_N$,
the sum $\sum'_{N'}$ avoiding the initial level $N$ \cite{pertur}.
For an excited state, $\int_{t_0}^t dt'\Delta E_N(t')=
\int_{t_0}^t dt' (E^L_n + \hbar A_n/2i) -\frac{1}{2}i\hbar \mathcal{N}_n$ with $\mathcal{N}_n$
a time-independent normalization of the wave function. Lamb shift and spontaneous emission rate are
\begin{equation}\label{lamb}
   E^L= \hbar \omega_c\frac{2 \alpha }{3\pi} \,  x\log \frac{2}{x} \ \ ,\ \
   A_n=  n\omega_c\frac{4\alpha }{3}    x
\end{equation}
with $x = \hbar \omega_c/\mu c_0^2 $ and $\alpha = q^2/\hbar c_0$ the fine structure constant; $E^L$  is equal for all Landau levels
whereas radiative decay is proportional to rotational energy.

In reciprocal space, the transverse EM momentum $J_z$ in Eq.~(\ref{LLkinQV}) is expressed as
$ ({J}_z)(\mathbf{k})_{ij} = -i\hbar \delta_{ij} (\mathbf{k} \times \mathbf{\nabla}_\mathbf{k})_z - i\hbar\epsilon_{zij}$,
i.e. as a sum of orbital angular momentum and
spin \cite{Cohen}. Neither one of them behaves as a genuine angular momentum \cite{Enk} but this separation is physically useful. In Hilbert space $J_z$ reads,
\begin{equation}\label{Jzperp}
    J_z = \hbar\sum_\mathbf{k} \alpha_i^\dag(\mathbf{k})  { J}_{ij}(\mathbf{k})  \alpha_j(\mathbf{k})
\end{equation}
with the photon annihilation operator $ \alpha_i(\mathbf{k}) = k^{-1}  \sum_{\mathbf{\Pi}} a_{k\mathbf{\Pi}} {\Phi}_{\mathbf{\Pi},i }(\mathbf{\hat{k}})$, and its associated  creation operator
$ \alpha^\dag_i(\mathbf{k}) $. The quantum expectation of $J_z$  is obtained by inserting the linearly perturbed eigenfunction~(\ref{perturb}) on both
sides of
the matrix element $\langle \Psi_{N}(t)|  J_z | \Psi_{N}(t)\rangle$. This creates either a virtual or real photon with energy $\hbar \omega_k$ and angular momentum
$\mathbf{\Pi}''$ out of the quantum vacuum at the cost of canonical angular momentum of the charge. Working out the photon operators,
leaves us with
\begin{eqnarray*}
  \langle J_z  \rangle &=& \frac{2 \pi \hbar q^2}{\mu^2 c_0} \mathrm{e}^{2\epsilon t}   \sum_{\mathbf{k}} \frac{1}{k^2}\sum_{\mathbf{k'}\mathbf{\Pi}'}
  \sum_{\mathbf{k}''\mathbf{\Pi}''}  \frac{\delta_{kk'} \delta_{k''k}}{(k')^{3/2}(k'')^{3/2} } \\
  &\times & \langle n| \mathbf{p}^K\cdot \mathbf{\Phi}_{\mathbf{\Pi'}}(\hat{\mathbf{k}}') \mathrm{e}^{i\mathbf{k'r}}
        \frac{1}{E_n-H_0' -\hbar \omega_{k} -i\epsilon } \\
  & \times &\bar{\mathbf{\Phi}}_{\mathbf{\Pi' }i}(\hat{\mathbf{k}}) { J}_{ij}(\mathbf{\mathbf{k}})
   {\mathbf{\Phi}}_{\mathbf{\Pi''} j}(\hat{\mathbf{k}}) \\
&\times&      \frac{1}{E_n-H_0' -\hbar \omega_{k} +i\epsilon }  \mathbf{p}^K\cdot
\bar{\mathbf{\Phi}}_{\mathbf{\Pi''}}(\hat{\mathbf{k}}'') \mathrm{e}^{-i\mathbf{k''r}} |n \rangle
\end{eqnarray*}
In principle is $\mathbf{\Pi}'=\mathbf{\Pi}''$ since the spherical harmonics are orthogonal eigenfunctions of  ${ J}_{ij}(\mathbf{k})$. The mathematics
is  easier
by using their completeness,  $    \sum_\mathbf{\Pi} \bar{\mathbf{\Phi}}_{\mathbf{\Pi}}(\hat{\mathbf{k}}){\mathbf{\Phi}}_{\mathbf{\Pi}}(\hat{\mathbf{k}}') =
    \delta_{\mathbf{\hat{k}}\hat{\mathbf{k}}'}\Delta(\mathbf{\hat{k}})$,
with $\Delta(\mathbf{\hat{k}})$  transverse  to $\mathbf{\hat{k}}$ imposed by the Coulomb gauge. The exponential $\exp(i\mathbf{k\cdot r})$
can be moved by using the operator identity $
\exp(i\mathbf{k\cdot r}) f(\mathbf{p}) = f(\mathbf{p}-\hbar \mathbf{k})\exp(i\mathbf{k\cdot r})$, and which induces a photon recoil in $H_0(\mathbf{p})$.
With $\delta_{\mathbf{kk}'} =
\delta_{\mathbf{\hat{k}\hat{k}}'} \delta_{{kk}'} /k^2 $,
 \begin{eqnarray*}
\langle J_z \rangle &=& \frac{2 \pi \hbar \mathrm{e}^{2\epsilon t/\hbar} q^2}{\mu^2 c_0}   \sum_{\mathbf{k}\mathbf{k}''}
  \langle n| {p}_m^K \frac{1}{E_n-H_0'(\mathbf{p}-\hbar \mathbf{k}) -\hbar \omega_{k} -i\epsilon }\\
  &&      \ \ \  \frac{1}{k}\mathrm{e}^{i\mathbf{k'r}} \Delta_{mi} (\mathbf{\hat{k}}) {{J}}_{ij}(\mathbf{\mathbf{k}}) \Delta_{jl}(\mathbf{\hat{k}}'') \mathrm{e}^{-i\mathbf{k''r}}
  \delta_{\mathbf{k}\mathbf{k}"}   \\
&&   \frac{1}{E_n-H_0'(\mathbf{p}-\hbar \mathbf{k}'') -\hbar \omega_{k} +i\epsilon } {p}_l^K |n \rangle
\end{eqnarray*}
From  this expression the EM spin of the quantum vacuum can be identified as,
 \begin{eqnarray}\label{S}
\langle S_z\rangle &=& \frac{1}{3}\frac{2 \pi \hbar \mathrm{e}^{2\epsilon t/\hbar} q^2}{\mu^2 c_0}   \frac{\hbar}{i}\epsilon_{zij} \times\nonumber  \\
  && \sum_{\mathbf{k}} \frac{1}{k}\langle n| {p}_i^K \frac{1}{|E_n-H_0' -\mathcal{E}(k) +i\epsilon |^2} {p}_j^K |n \rangle
\end{eqnarray}
We have performed the angular integral over $\mathbf{k}$ to eliminate $\Delta(\mathbf{\hat{k}})$, neglected the photon recoil
$\mathbf{p}^K\cdot \hbar \mathbf{k}/\mu$ irrelevant for spin, and defined the energy $\mathcal{E}(k) = \hbar \omega_k + \hbar^2k^2/2\mu$.
The orbital angular momentum is associated with a
 differential operator acting on $ \delta_{\mathbf{k}\mathbf{k}"}$, and an integration by parts is imposed to perform
the integral over $\mathbf{k}''$. Since this operator acts only on angles, we find for the orbital angular momentum,
 \begin{eqnarray}\label{L}
\nonumber \langle L_z\rangle &&= \frac{2 \pi \hbar \mathrm{e}^{2\epsilon t/\hbar} q^2}{\mu^2 c_0}   \sum_{\mathbf{k}}
  \langle n| {p}_m^K \times \frac{\hbar}{i}\epsilon_{zst} \\
 \nonumber  &&\hat{ k}_t\nabla_s\left( \Delta_{mi}(\mathbf{\hat{k}})
  \frac{1}{E_n-H_0'(\mathbf{p}-\hbar \mathbf{k}) -\hbar \omega_{k} -i\epsilon }\mathrm{e}^{i\mathbf{kr}}  \right)
  \\
  &&     \Delta_{il}(\mathbf{\hat{k}}) \mathrm{e}^{-i\mathbf{kr}}
\frac{1}{E_n-H_0'(\mathbf{p}-\hbar \mathbf{k}) -\hbar \omega_{k} +i\epsilon } {p}_l^K |n \rangle
\end{eqnarray}
As $k_z=0$ the kinetic operator $\mathbf{p}^K$ is located in the $xy$ plane. It is customary to write $p^K_x +(-) i p^K_y = (2\mu \hbar \omega_c)^{1/2}
c^{(\dag)}$ in terms
of the raising and lowering
operators of the Landau levels, in terms of which $H_0 = \hbar \omega_c (c^\dag c + \frac{1}{2})$. To evaluate the spin in Eq.~(\ref{S}) we use
$\epsilon_{zij} p^K_i f(H_0) p^K_j=
-i\mu \hbar \omega_c (c^\dag f(E_{n-1})c -  c f(E_{n+1})c^\dag)$. The first term  implies the release of a real photon with energy $\hbar \omega_c$.
As $\epsilon \downarrow 0$,
this part of $\langle S_z\rangle$ is written as   $(\hbar/2)\int_{t_0}^tdt' \exp(2\epsilon t'/\hbar) \delta(\hbar\omega_c -\hbar \omega_k)$, so that
\begin{equation}\label{dSdt}
    \frac{d}{dt}\langle S_z\rangle = -\frac{1}{2} A_n \hbar
\end{equation}
with $A_n$ defined in Eq.~(\ref{lamb}). The second term involves virtual photons and is finite as $\epsilon \downarrow 0$,
\begin{eqnarray}\label{Sn}
    \langle S_z\rangle &=& (n+1) \frac{q^2\hbar^3 \omega_c}{3\mu c_0}  \int_0^\infty  \frac{k dk}{(\hbar\omega_c + \mathcal{E}(k) )^2} \nonumber \\
    &=&\alpha \frac{\hbar}{3\pi} (n+1) \, x \log \frac{2}{x}
\end{eqnarray}
In expression~(\ref{L}) for  $\langle L_z\rangle$ the derivative $\nabla_\mathbf{k}$ acts on 3 factors inside brackets. Its action on the
factor in the middle, caused by the photon recoil,
is a factor $x$ smaller than the rest. The action on the first factor gives $\epsilon_{zst} \hat{k}_t\nabla_s(\Delta_{mi} ) \Delta_{il} =  -k^{-1}
\epsilon_{zlt}\hat{k}_m\hat{k}_t$ and produces an angular momentum
$\langle L^{(1)}_z\rangle=\langle S_z\rangle $. Finally, the action of $\nabla_\mathbf{k}$ on
$\exp(i\mathbf{k}\cdot \mathbf{r}) $ leads to the expression
\begin{eqnarray*}
\langle L^{(2)}_z\rangle &=& \frac{2 \pi \hbar^2 \mathrm{e}^{2\epsilon t/\hbar} q^2}{\mu^2 c_0} \times \\
   &&  \sum_\mathbf{k}  \langle n| p^K_m \Delta_{ml}(\mathbf{\hat{k}}) \frac{1}{\mathcal{H}_n -i\epsilon}
(\mathbf{r}\times \mathbf{\hat{k}})_z \frac{1}{\mathcal{H}_n + i\epsilon} p^K_l |n\rangle
\end{eqnarray*}
where $\mathcal{H}_n \equiv E_n - H_0 - \mathcal{E}(k) + {\hbar\mathbf{ p}^K\cdot \mathbf{k}}/{\mu}$.
This time, the photon recoil cannot be ignored and we must expand either one of the denominators
which produces an
integral $dk \hat{d\mathbf{k}}$ with integrand of the type
\begin{equation*}
    \epsilon_{zst} \Delta_{ml}(\mathbf{\hat{k}})k_u \hat{k}_t \times p^K_m \frac{1}{\mathcal{H}_n - i\epsilon} p^K_u \frac{1}{\mathcal{H}_n - i\epsilon}  r_s  \frac{1}{\mathcal{H}_n + i\epsilon}
p^K_l
\end{equation*}
Physically, this corresponds to the creation of a virtual photonic mode with finite orbital angular momentum.
In terms of the infinitely degenerated center $(X,Y)$ of the cyclotron orbit we associate $ x = X-p_y^K/\mu \omega_c$ and  $ y = Y +p_x^K/\mu \omega_c $. The operators $(X,Y)$ drop out in the vacuum expectation
 value since
they do not occur in pairs. Upon expressing $(p^K_x,p^K_y)$  in the operators $c$ and $c^\dag$, the integrand contains four transition operators.
As was the case for $ \langle S_z\rangle $, some contribute to spontaneous emission but are seen to be a factor $x$ smaller than
$A_n$. We thus focus on terms where
the limit $\epsilon \downarrow 0$ exists. For instance, the sequence
\begin{equation*}
   \langle n| c\mathcal{H}_n^{-1} c^\dag \mathcal{H}_n^{-1} c \mathcal{H}_n^{-1}  c^\dag |n\rangle = \frac{(n+1)(n+2) }{\mathcal{E}(k)^3}
\end{equation*}
leads to,
\begin{eqnarray*}
     \langle L^{(2)}_z\rangle &\sim &  (n+1)(n+2) \frac{2 \pi \hbar^2 q^2}{\mu^2 c_0} \frac{\hbar}{\mu} \mu \hbar^2 \omega_c
     \sum_\mathbf{k}\frac{k}{\mathcal{E}(k)^3} \\
&\approx& \frac{\hbar}{8\pi }\alpha (n+1)(n+2) \, x
\end{eqnarray*}
Upon collecting all possibilities, performing the angular integral, and adding the complex conjugate,
we obtain\begin{eqnarray}\label{dJdt}
             \langle J_z\rangle = \langle S_z\rangle -\frac{4\hbar}{15\pi } \alpha (n+1)(n+4)\, x
         \end{eqnarray}
and ${d}/{dt}  \langle J_z\rangle  = -\hbar A_n/2$.

The last but one term in Eq.~(\ref{LLkinQV}), the longitudinal EM angular momentum $\Delta L_\parallel$, is itself linear in the vacuum field. Its leading quantum
expectation value is obtained using the linear perturbation of the eigenfunctions and  the completeness of the spherical vector harmonics,
\begin{eqnarray*}
  \langle \Psi_N| \Delta L_\parallel | \Psi_N\rangle  = -\frac{q^2}{\mu c_0} \frac{4\pi \hbar}{3} \sum_\mathbf{k} \frac{1}{k}
\epsilon_{zij} \langle n |r_i \frac{1}{\mathcal{H}_n+ i\epsilon } p^K_j|n \rangle
\end{eqnarray*}
plus its $c.c.$ Here, the $k$-integral diverges in the UV as $dk/k$. We
recognize for large $k$ the form $ (\delta \mu/ \mu) \langle n | l^K_z| n\rangle$, with $\delta \mu$ the well-known Bethe-Kramers mass
renormalization \cite{milonni}. Upon adding it to the kinetic mass as $\mathbf{l}^K = (\mu+\delta\mu) \mathbf{r} \times d\mathbf{r}/dt$,
and upon subtracting it from the above equation, we obtain,
\begin{eqnarray*}
  &&\langle \Psi_N| \Delta L_\parallel | \Psi_N\rangle  = -\frac{q^2}{\mu c_0} \frac{4\pi \hbar}{3} \sum_\mathbf{k} \frac{1}{k} \sum'_{n'} \frac{i\hbar}{\mu} \times \\
  && \epsilon_{ijz} \frac{\langle n |p^K_i |n'\rangle}{\Delta E_{n'n}   + \mathcal{E}(k) -i\epsilon } \frac{\langle n' |p^K_j|n
  \rangle}{\mathcal{E}(k)}+ c.c. \\
\end{eqnarray*}
where we have used the identity  $ \Delta E_{n'n} \langle n |\mathbf{r} | n'\rangle  =  (i\hbar/\mu)  \langle n | \mathbf{p}^K| n'\rangle $.
The operators $p^K_x$ and $p^K_y$ can be expressed
in terms of $c^\dag,\, c$, which results in,
\begin{equation}\label{Lpara}
\langle  \Delta L_\parallel  \rangle   = \hbar \frac{4 \alpha }{3\pi} \, x \log \frac{2}{x}
\end{equation}
which, like the Lamb shift in energy, is independent on $n$.

The last contribution of the quantum vacuum to the angular momentum is associated with the Lenz term $l^T$
in Eq.~(\ref{LLkinQV}). The quantum vacuum comes in via $\mathcal{{N}}_n$  and
the second term in Eq.~(\ref{perturb}).
With $ l^T = \hbar( c^\dag c+ 1 + b^\dag b -ib^\dag c^\dag/2 + icb/2 ) $ in terms of the raising ($b^\dag$) and lowering ($b$) operator of the
degenerated $m$-levels \cite{Goerbig}, the $m$-dependence is seen to cancel in the sum of both terms. The second term equals
$\hbar(n+1)$ times
\begin{eqnarray*}
    \frac{4\pi \hbar q^2}{3\mu^2 c_0} \frac{1}{i\hbar} \int_{t_0}^t \, dt' \mathrm{e}^{2\epsilon t'/\hbar}
\sum_\mathbf{k}\langle n |  {p}^K_j  \frac{1}{\mathcal{H}_n +i\epsilon}  {p}^K_j  |n\rangle + c.c. \\
  = - A_n \int_{t_0}^t \, dt' \mathrm{e}^{2\epsilon t'/\hbar} -\frac{2\alpha}{3\pi }(n+1) \,  x \log  \frac{2}{x}
\end{eqnarray*}
and the first is $\hbar$ times
\begin{eqnarray*}
    \frac{4\pi \hbar q^2}{3\mu^2 c_0}  \mathrm{e}^{2\epsilon t/\hbar}
\sum_\mathbf{k}\frac{1}{k}\langle n |{p}^K_j  \frac{1}{\mathcal{H}_n +i\epsilon}
(c^\dag c +1)  \frac{1}{\mathcal{H}_n -i\epsilon}  {p}^K_j |n\rangle \\
= n  A_n \int_{t_0}^t \, dt' \mathrm{e}^{2\epsilon t'/\hbar} +
\frac{2\alpha }{3\pi }(n+1)(n+2)  x \log  \frac{2}{x}
\end{eqnarray*}
In particular, $n^2$ terms also cancel and
\begin{equation}\label{Barnett}
    \langle \delta l^T \rangle = 2 \langle S_z \rangle\, ;  \frac{d}{dt}\langle \delta l^T \rangle  = - \hbar A_n
\end{equation}
By adding up the four contributions $\langle S_z\rangle$, $\langle J_z\rangle$, $\langle J_\parallel\rangle $ and
$\langle \delta l^T\rangle$ we find for the total angular momentum of the quantum vacuum
\begin{eqnarray}\label{total}
\langle J_z^{QV} \rangle = \hbar \frac{4\alpha}{3\pi }(n+2)\,  x \log  \frac{2}{x} -\hbar\frac{4\alpha}{15\pi }  (n+1)(n+4) x \nonumber \\
\end{eqnarray}
and $ {d}\langle J_z^{QV}\rangle/dt   = -2\hbar A_n$.
We conclude that the quantum vacuum achieves an angular momentum that is, in units of $\hbar$, proportional to
$ \alpha \times x= \alpha \times
\hbar \omega_c/\mu c_0^2 \approx    10^{-12}/$Tesla, which is time-dependent if $B_0$ is.
In all momentum integrals the photon momentum $\hbar k $ takes values up to
$\mu c_0$ with nonetheless a significant weight of non-relativistic momenta. A relativistic description of the rotating electron is thus relevant but
should affect
only numerical coefficients in Eq.~(\ref{total}).
Even in a relativistic picture,
Eq.~(\ref{LLkinQV}) for the kinetic
angular momentum is valid, and $\langle l^K_z \rangle $ remains quantized to $-(2n+1)\hbar$ \cite{noteNR}. In the ground state, the existence of Casimir angular momentum
makes the kinetic angular momentum
slightly more negative than $-\hbar$, in states with large $n$ it will be slightly less negative than $-(2n+1)\hbar $,
the correction growing like $n^2$. Note that the gauge-invariant magnetic moment
$M_z= (q/2\mu) l^K$ of the rotating charge is subject to the same correction.
Due to the quantum vacuum, the kinetic angular $-(2n+1) \hbar$ decays to the Landau level $n-1$ with rate $A_n$ so that
$d\langle l^K \rangle/dt = + 2\hbar A_n$ and $\langle l_z^K+ J_z^{QV} \rangle =\mathcal{ L}_z$ is conserved in the decay.
 The non-relativistic analysis imposes that
$E_n \ll \mu c_0^2$, implying, for an electron in a field of 10 Tesla, that $ n \ll 10^9$. Pushing our theory to this extreme synchrotron regime,
the relative contribution
of Casimir orbital angular momentum would be
of order $10^{-4}$.

\section{Conclusions}
The main objective of this work is to establish the existence of  angular momentum of the EM quantum
vacuum, induced by the presence of a rotating charge in a magnetic field. It is instructive to look at the separate contributions of
 spin, orbital angular momentum and angular momentum directly associated with the gauge fields. All are
oriented along the magnetic field and proportional to the product of fine
structure constant and the small ratio of
rotational energy to rest energy.
Spin and orbital angular momentum  decay in the sam way, their coupling being large, yet
the orbital angular momentum of the quantum vacuum, induced by photon recoil, dominates angular momentum for highly
energetic Landau levels.
Casimir angular momentum is governed by virtual photons with energies up to the
rest mass of the charge and, despite the UV renormalizability of the theory, would merit a relativistic treatment.
A future challenge would be to study EM angular momentum in the fully relativistic  synchrotron problem, or to investigate it for
Rydberg orbits.


\begin{thebibliography}{4}
\bibitem{pertur} We ignored the perturbation  $|\delta \Psi_{N}\rangle = \sum'_{N'N''} |N'\rangle {W_{N'N''}W_{N''N}}/({(E_{NN'} +2i\epsilon) (E_{NN''} +i\epsilon) })$.
For one-photon processes, the  matrix element $\langle n,\{0\} | l^K_z|\delta \Psi_{n\{0\}}\rangle $ vanishes.
\bibitem{noteNR} The eigenfunctions are the same for the relativistic Hamiltonian without spin, and thus the Lenz moment $\langle l^T \rangle $ remains unaltered,
that in turn determines $l^K_z$\cite{Cohen}.
\end{thebibliography}

\end{document}